\documentclass[preprintnumbers,nofootinbib,showpacs,eqsecnum,pre,12pt]{revtex4-1}
\usepackage{amsmath}         
\usepackage{amssymb}          
\usepackage[pdftex]{graphicx}	
\usepackage{grffile}
\usepackage[export]{adjustbox}
\usepackage[pdftex]{hyperref} 
\usepackage[utf8x]{inputenc}
\usepackage{dsfont}
\usepackage{array,multirow}
\usepackage{ulem}
\usepackage{xspace}
\usepackage{enumitem}
\usepackage{nccmath}
\usepackage{afterpage}

\newcommand{\ord}[1]{\mathcal{O}\left({#1}\right)}

\def\gsim{\raise0.3ex\hbox{$\;>$\kern-0.75em\raise-1.1ex\hbox{$\sim\;$}}}
\def\lsim{\raise0.3ex\hbox{$\;<$\kern-0.75em\raise-1.1ex\hbox{$\sim\;$}}}

\begin{document}

\title{Search for Long-Lived Heavy Neutrinos at the LHC with a VBF Trigger}

\author{J. Jones-P\'erez}
\email[E-mail: ]{jones.j@pucp.edu.pe}
\author{J. Masias}
\email[E-mail: ]{j.masias@pucp.pe}
\affiliation{
Secci\'on F\'isica, Departamento de Ciencias, Pontificia Universidad Cat\'olica del Per\'u, Apartado 1761, Lima, Peru
}
\author{J. D. Ruiz-\'Alvarez}
\email[E-mail: ]{josed.ruiz@udea.edu.co}
\affiliation{
Instituto de F\'isica, Universidad de Antioquia, A.A. 1226 Medell\'in, Colombia.
}

\begin{abstract}

The charged current production of long-lived heavy neutrinos at the LHC can use a prompt charged lepton for triggering the measurement of the process. However, in order to fully characterize the heavy neutrino interactions, it is necessary to also probe Higgs or $Z$ mediated neutral current production. In this case the charged lepton is not available, so other means of triggering are required.

In this work, we explore the possibility of using a vector boson fusion trigger in the context of a GeV-scale Type I Seesaw model. We consider a minimal model, where both Higgs and $Z$-mediated contributions produce one heavy neutrino, as well as an extended model where the Higgs can decay into two heavy ones. Both scenarios are tested through displaced dilepton and displaced multitrack jet searches.

\end{abstract}

\maketitle

\section{Introduction}
\label{sec:intro}

The discovery of non-zero neutrino masses strongly suggests that these particles are the surest window towards physics beyond the Standard Model (SM). Out of the several models explaining neutrino mass, the Type-I Seesaw~\cite{Minkowski:1977sc,GellMann:1980vs,Yanagida:1979as,Mohapatra:1979ia} is probably the most popular one. Apart from giving a Majorana mass to the light SM neutrinos, its main prediction is the existence of heavy neutrinos, also referred to as heavy neutral leptons, which couple through charged ($W^\pm$ mediated) and neutral ($Z$ and Higgs mediated) interactions. 

Producing these heavy neutrinos at a collider is not expected to be easy. The naive seesaw prediction is that, if the heavy neutrinos have masses small enough for them to be generated at a current collider, their coupling with other SM particles will be so small that the probability of them ever being produced will effectively vanish. Nevertheless, it is well known that the presence of a lepton number-like symmetry can allow accessible masses with relatively large mixing, respecting at the same time the smallness of SM neutrino masses~\cite{Mohapatra:1986bd,Shaposhnikov:2006nn,Kersten:2007vk,Gavela:2009cd,Ibarra:2010xw,LopezPavon:2012zg,Antusch:2015gjw,Hernandez:2018cgc}. This fact has encouraged several experimental searches for them, which have been reviewed, for example, in~\cite{Atre:2009rg,Deppisch:2015qwa,Das:2018hph}.

A particularly interesting type of search is the one for long-lived heavy neutrinos. In fact, for small enough mixing, heavy neutrinos with masses under $\sim50\,$GeV can have macroscopic decay lengths~\cite{Gronau:1984ct}. In the past years, many works have proposed searches for such long-lived particles at the LHC~\cite{Helo:2013esa,Cui:2014twa,Izaguirre:2015pga,Gago:2015vma,Duarte:2016caz,Caputo:2017pit,Antusch:2017hhu,Abada:2018sfh,Cottin:2018kmq,Cottin:2018nms,Drewes:2019fou,Drewes:2019vjy,Das:2019fee,Chiang:2019ajm,Alimena:2019zri}. As with any search, these have always had to specify a trigger for the process, with most of the latter searches using a prompt charged lepton coming from the primary vertex. This has actually been used recently by the ATLAS collaboration to start constraining part of the relevant parameter space~\cite{Aad:2019kiz}.

Triggering with a prompt charged lepton is very likely to be the most reasonable way to search for, and discover, these kind of heavy neutrinos. Since the latter need to be relatively light to be long-lived, their decay products are somewhat soft, and triggering with them is not very efficient. Moreover, by detecting the prompt lepton one can be sure that heavy neutrino production has taken place through a \textit{charged current} coupling, that is, through a coupling with a $W^\pm$ boson.

However, once a particle is discovered it is desirable to measure as much of its properties as possible, similar to what has been done e.g.\ with the Higgs boson~\cite{Aad:2019mbh, Sirunyan:2018sgc}. Thus, one can ask if heavy neutrino production via a \textit{neutral current} is measurable at the LHC, such as by the decay of a $Z$ or Higgs boson. In particular, couplings with the Higgs can only be probed by production processes, as decays with intermediate $W^\pm$ and $Z$ have much higher branching ratios. In any case, in neutral current production the heavy neutrino is generated along a light SM neutrino, meaning that the prompt charged lepton trigger is no longer available. Furthermore, the energy of the outgoing SM neutrino is not large enough to be used by the missing energy trigger. This usually leaves the heavy neutrino soft decay products as the only candidates for triggering. For example, in~\cite{Gago:2015vma} the decay of Higgs bosons into heavy neutrinos was studied, with the former being produced by gluon fusion. It was found that even though the branching ratios could be large enough to produce hundreds of events at the LHC, using the standard triggers would cause the efficiency to drop drastically.

In this paper, we propose using a vector boson fusion (VBF) trigger~\cite{Khachatryan:2015bnx,ATL-DAQ-PUB-2019-001,CMS-DP-2018-005} to explore neutral current production of long-lived heavy neutrinos. By using VBF we stop depending on the soft final states for triggering, relying instead on particles participating on the production process. In principle, the VBF cross-section is much smaller than that of $W^\pm$-mediated Drell-Yann, so this search would not be expected to serve as a discovery channel. However, as we will see, in the presence of effective operators, VBF provides a way of probing regions of the parameter space with very large masses and small mixing, usually inaccessible to other production channels.

To this end, in Section~\ref{sec:seesaw} we provide a brief overview of the models of interest. In Section~\ref{sec:vbf} we give details on the VBF trigger and describe the two different final states we use to probe the models. In Section~\ref{sec:results} we present the results of our simulation. In addition, on the Appendix we show our parametrization, and give simple formulae for several limit cases.

\section{The Type-I Seesaw}
\label{sec:seesaw}
\subsection{Minimal Model}

The Type-I Seesaw~\cite{Minkowski:1977sc,GellMann:1980vs,Yanagida:1979as,Mohapatra:1979ia} explains the smallness of SM neutrino masses by adding heavy Majorana fermions $\nu_R$, singlets under the SM, which we refer to as \textit{sterile} neutrinos. The Lagrangian is extended by the following terms:
\begin{equation}
\mathcal L = \mathcal L_{SM}-\bar L_a\, (Y_\nu)_{as}\, \nu_{R_s} \tilde \phi-
\frac{1}{2}\bar\nu_{R_s}^c (M_R)_{st}\, \nu_{R_t} + {\rm h.c.}
\end{equation}
where the index $a$ refers to the \textit{active} flavours: $e,\,\mu,\,\tau$. In this study we work with the $3+3$ Seesaw, meaning that we add three sterile neutrinos, with indices $s,t=1,\,2,\,3$. After electroweak symmetry breaking, the $6\times6$ neutrino mass matrix acquires the form:
\begin{equation}
M_\nu=\begin{pmatrix}
0 & m_D \\
m_D^T & M_R
\end{pmatrix}
\end{equation}
with $(m_D)_{as}=v\,(Y_\nu)_{as}/\sqrt2$.

After diagonalization by a $6\times6$ unitary matrix $U$, we obtain six mass eigenstates: $\nu_1,\nu_2,\,\nu_3,\,N_4,\,N_5,\,N_6$. Naively, one expects the active-heavy mixing $U_{ah}$ to be suppressed by $\sqrt{m_\ell/M_h}$, where $m_\ell$ ($M_h$) represent the light (heavy) neutrino masses. However, it is well known that, once one adds more than one sterile neutrino, it is possible to enhance the mixing~\cite{Casas:2001sr,Donini:2012tt}. This is shown in further detail on the Appendix, and is attributed to the presence of a softly broken lepton number symmetry~\cite{Mohapatra:1986bd,Kersten:2007vk,Gavela:2009cd,Ibarra:2010xw,LopezPavon:2012zg,Antusch:2015gjw,Hernandez:2018cgc}.

For the $3+2$ Seesaw, once the parameters in the active-light sector are fixed, the resulting structure is very rigid. For normal ordering (NO) and CP conservation, the heavy neutrinos mix preferably with $\mu,\tau$ flavours, having an additional suppression of order $s_{13}$ for the mixing with $e$ flavour~\cite{Donini:2012tt,Gago:2015vma}. For instance, for $m_1=0.01$~eV, we have:
\begin{equation}
|U_{e4}|^2:|U_{\mu4}|^2:|U_{\tau4}|^2=0.08:0.50:0.42
\end{equation}

In contrast, for the $3+3$ case, since more parameters are available, it is possible to avoid this constraint and have mixings with all flavours on a similar footing. Details of possible flavour patterns can be found on the Appendix\footnote{Notice that it is also possible to obtain different patterns by varying $\delta_{CP}$ and the Majorana phases in $U_{\rm PMNS}$~\cite{Asaka:2011pb,Chrzaszcz:2019inj}}. For this work, we choose NO with option~\ref{item:No-s13-1}:
\begin{eqnarray}
U_{a4}
&=& z_{45}\,Z_a\,\sqrt{\frac{m_2}{M_4}}\cosh\gamma_{45}\,e^{i\,z_{45}\,\theta_{45}} \\
U_{a5}
&=& i\,Z_a\sqrt{\frac{m_2}{M_5}}\cosh\gamma_{45}\,e^{i\,z_{45}\,\theta_{45}} \\
U_{a6}
&=& i\,(U_{\rm PMNS})_{a3}\sqrt{\frac{m_3}{M_6}}
\end{eqnarray}
where $m_i$ ($M_j$) is the mass of $\nu_i$ ($N_j$), $\theta_{45}$ and $\gamma_{45}$ are the real and imaginary parts of a complex angle, respectively, $z_{45}$ is the sign of $\gamma_{45}$, and:
\begin{equation}
Z_a = (U_{\rm PMNS})_{a2}+i\,z_{45}\,\sqrt{\frac{m_1}{m_2}}(U_{\rm PMNS})_{a1}
\end{equation}
Given the unenhanced mixing of $N_6$, we see that if we take a large $M_6$ this scenario becomes an effective $3+2$ Seesaw, with two heavy neutrinos having enhanced active-heavy mixing. However, in this case the mixing with the $e$ flavour is largest. For the same $m_1$, we have:
\begin{equation}
|U_{e4}|^2:|U_{\mu4}|^2:|U_{\tau4}|^2=0.46:0.22:0.32
\end{equation}
This choice allows us to somewhat decrease decays into taus, which are harder to probe at colliders.

In this model, the production and decay of a heavy neutrino are essentially ruled by $|U_{ah}|^2$. This means that the small decay widths of long-lived particles are correlated to tiny production cross-sections. This can be a problem for $M_h\gtrsim40\,$GeV, which requires $|U_{ah}|^2\lesssim10^{-7}$ to be long-lived~\cite{Helo:2013esa,Gago:2015vma,Drewes:2019fou}, such that production cross-sections become too small to generate a signal at a collider.

In the following we set $M_4=M_5$, which reduces both the heavy neutrino contribution to neutrinoless double beta decay, and loop corrections to the light masses~\cite{LopezPavon:2012zg,Gago:2015vma,Lopez-Pavon:2015cga,Hernandez:2018cgc}.

\subsection{Extended Model}

After adding the new $\nu_R$, one can extend the model by including effective operators involving the new states~\cite{Graesser:2007yj,delAguila:2008ir,Aparici:2009fh,Caputo:2017pit,Liao:2016qyd,Butterworth:2019iff}. One possibility, which could arise from a $U(1)_{B-L}$ extension~\cite{Accomando:2016rpc}, involves two heavy neutrinos and two Higgs doublets:
\begin{equation}
\Delta\mathcal{L}=-\frac{\lambda_{st}}{\Lambda}\,\bar\nu^c_{R_s}\,\nu_{R_t}\,\phi^\dagger\,\phi+{\rm h.c.}
\end{equation}
After electroweak symmetry breaking, this dimension-five operator can mediate the decay of the Higgs into two heavy neutrinos, provided they are light enough:
\begin{equation}
\mathcal{L}_{h^0\nu_h\nu_h}=-(\alpha_{NH})_{st}\,\bar\nu^c_{R_s}\,\nu_{R_t}\,h^0+{\rm h.c.}
\end{equation}
where $(\alpha_{NH})_{st}\equiv\lambda_{st}\,v_{\rm SM}/\Lambda$. In principle, this operator also leads to a new contribution to the heavy neutrino Majorana mass~\cite{Aparici:2009fh,Caputo:2017pit}, which we ignore\footnote{Notice that for large enough $\alpha_{NH}$, it might be necessary to tune $M_R$ to obtain small heavy neutrino masses.}.

The main advantage of introducing this operator lies on the separation between heavy neutrino production and decay, ruled by $\alpha_{NH}$ and $U_{ah}$, respectively. In this case one can have, for example, $M_4$ as large as $\sim60\,$GeV, with a small enough $|U_{a4}|^2$ to keep $N_4$ long-lived, without penalizing the $h^0\to N_4 N_4$ branching ratio. As we shall see, having two decaying heavy neutrinos also increases the overall branching ratio into final states of interest.

\subsection{Numerical Implementation}

In this work, the minimal model was implemented in \texttt{SARAH 4.14.0}~\cite{Staub:2008uz,Staub:2012pb,Staub:2013tta}. This generated code for \texttt{SPheno 4.0.3}~\cite{Porod:2003um,Porod:2011nf}, which calculated the heavy neutrino mass spectrum and branching ratios. We used \texttt{SSP 1.2.5}~\cite{Staub:2011dp} to carry out the parameter variation.

The effective operator for the extended model was added to the \texttt{HeavyN} model~\cite{Alva:2014gxa,Degrande:2016aje} in \texttt{FeynRules 2.3.33}~\cite{Christensen:2008py,Alloul:2013bka}. This allowed the calculation of the Higgs partial width to two heavy neutrinos using \texttt{MadWidth}~\cite{Alwall:2014bza}, which was later merged with the widths calculated by \texttt{SPheno} for the minimal model.

In both cases, the model was exported in the \texttt{UFO} format~\cite{Degrande:2011ua} into \texttt{MadGraph5\_aMC@NLO 2.6.7}~\cite{Alwall:2014hca}. After generating the events, the posterior hadronization is carried out by \texttt{PYTHIA 8.247}~\cite{Sjostrand:2006za}, which uses the \texttt{CTEQ66} PDF set \cite{Nadolsky:2008zw}. Depending on the analysis, the decay position of the heavy neutrino is computed from the MC truth data, or via the time of flight option in \texttt{MadGraph}.

\section{Searches with a Vector Boson Fusion Trigger}
\label{sec:vbf}

\begin{figure}[tb]
\begin{center}
\includegraphics[width=0.25\textwidth]{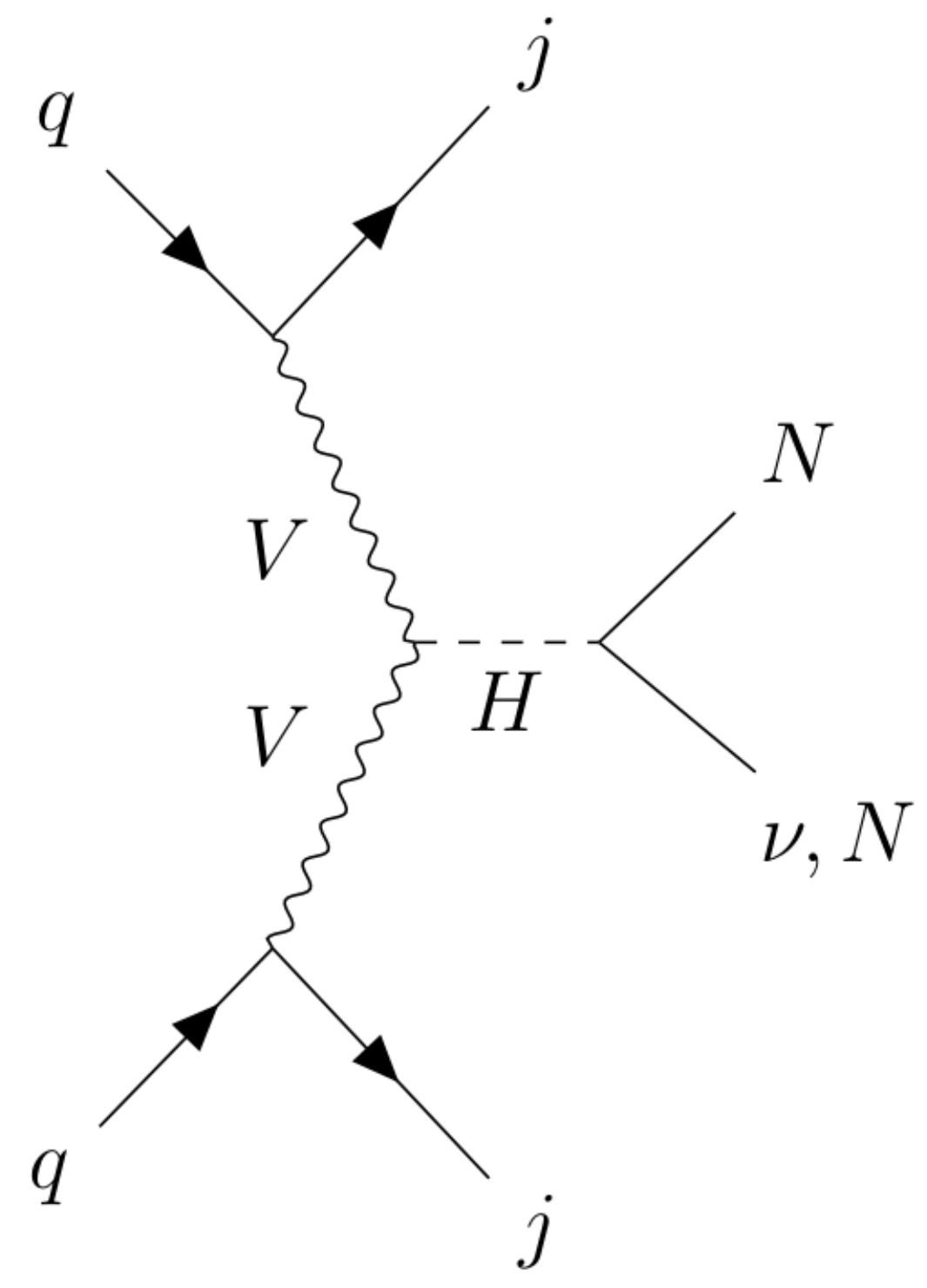} \qquad\quad
\includegraphics[width=0.25\textwidth]{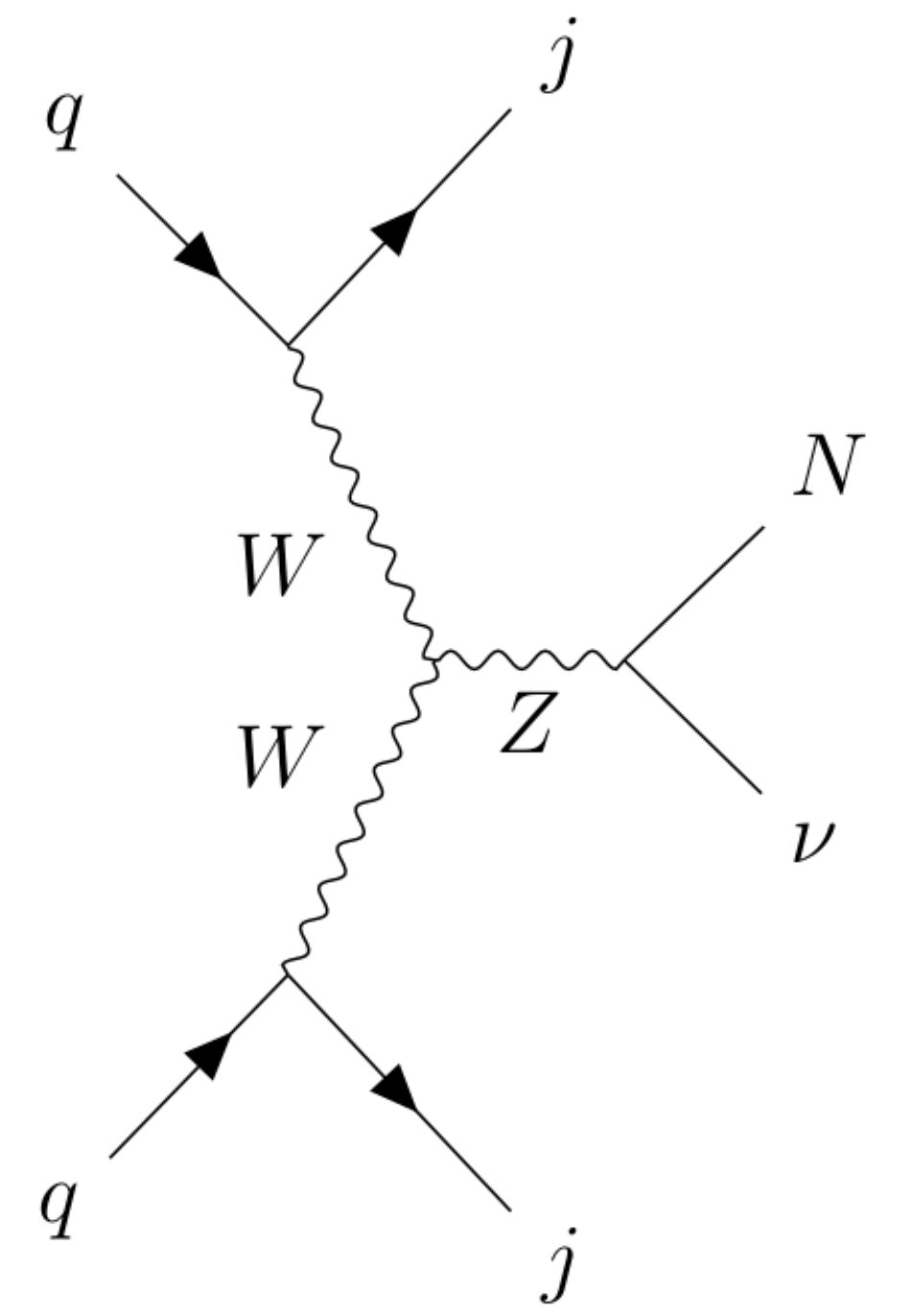} \qquad\quad
\includegraphics[width=0.20\textwidth]{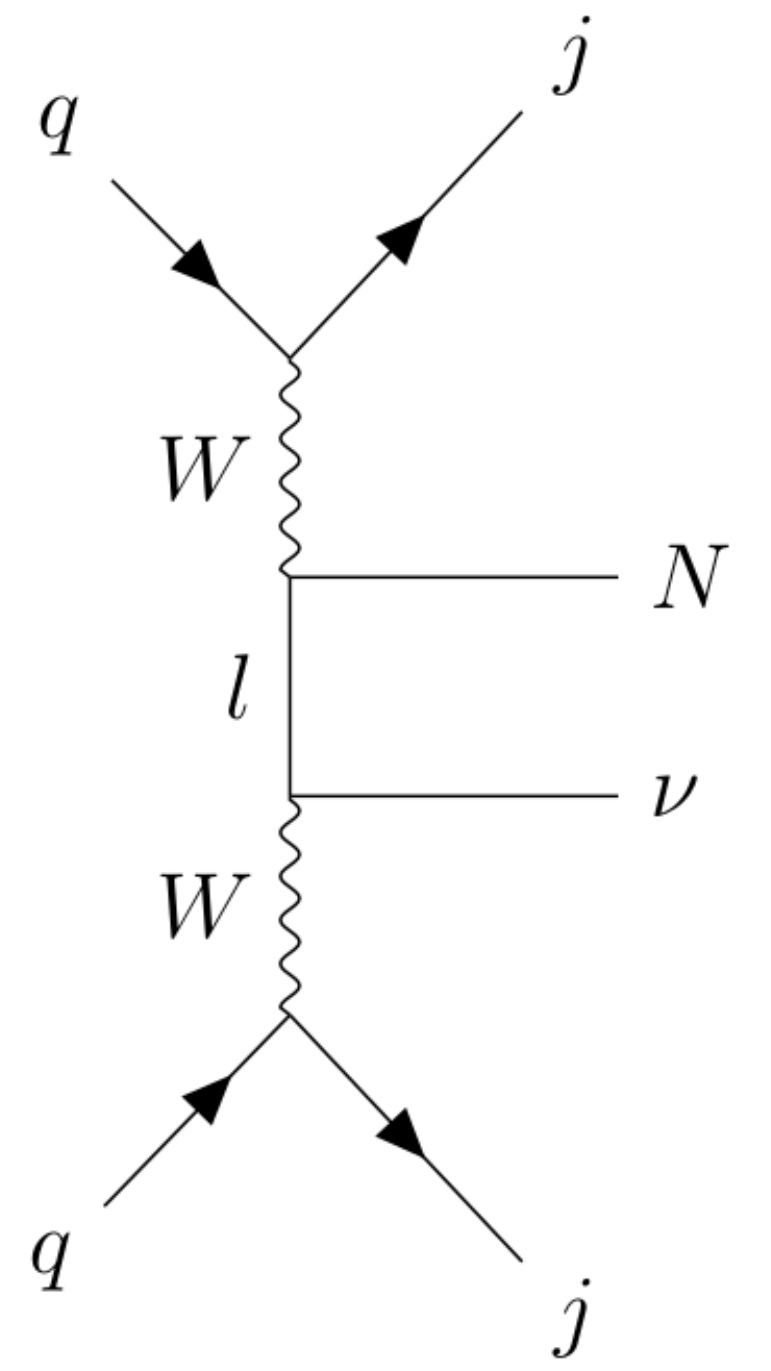}
\end{center}
 \caption{Different contributions to heavy neutrino production by VBF. $V$ denotes $W^\pm$ or $Z$ bosons}
 \label{fig:feynman1}
\end{figure}
In this work we propose to trigger the measurement of long-lived heavy neutrinos using the VBF production process~\cite{Cahn:1983ip,Dicus:1985zg,Altarelli:1987ue}. Such a trigger has already been used by the CMS experiment for Higgs boson studies~\cite{Khachatryan:2015bnx}, and is available at ATLAS~\cite{ATL-DAQ-PUB-2019-001}. As illustrated on the left and center diagrams of Figure~\ref{fig:feynman1}, our objective is to use VBF to produce either a $Z$ or Higgs boson, which will later decay into final states with heavy neutrinos. We have confirmed numerically that, in the region of the parameter space of interest, direct production through a pair of off-shell $W^\pm$, illustrated on the right diagram of Figure~\ref{fig:feynman1}, is around two orders of magnitude smaller than the former contributions. This certifies that VBF production effectively tests neutral current interactions.

\begin{table}[tb]
\centering
\setlength{\tabcolsep}{2em}
\begin{tabular}{| c | c || c | c |} 
\hline
$p_T(j_1)$& $>30$ GeV & $\eta(j_1)\cdot\eta(j_2)$& $<0$ \\
\hline
$|\,\eta(j_1)\,|$& $< 5.0$ & $|\,\Delta\eta(j_1, j_2)\,|$ & $>4.2$\\
\hline
$p_T(j_2)$& $>30$ GeV &  $m_{j_1 j_2}$ & $>750$ GeV\\
\hline
$|\,\eta(j_2)\,|$& $< 5.0$ & $\sum_j p_T$ & $>200\,$GeV \\
\hline
\end{tabular}
\caption{VBF selection criteria, where $j_1$ and $j_2$ refer to the two most energetic jets, and the sum is over all hadronic activity.}
\label{table:1}
\end{table}
The VBF topology is particularly useful to search for new physics, see for instance~\cite{Dutta:2012xe, Delannoy:2013ata,Liu:2014lda,Bambhaniya:2015wna,Andres:2016xbe,Andres:2017daw,Dutta:2017lny,Florez:2018ojp,Florez:2019tqr}. It involves two high $p_T$ jets in the forward direction, located in opposite hemispheres of the detector. The jets must have a large difference in pseudorapidity $|\Delta\eta|$, as well as a large invariant mass $m_{j_1j_2}$. In addition, the whole event must pass a cut over all hadronic activity. The event selection criteria are presented in Table~\ref{table:1}. It is important to take into account that other VBF cuts are also available in the literature~\cite{Khachatryan:2015bnx,ATL-DAQ-PUB-2019-001,CMS-DP-2018-005}, and we have checked that they do not modify our results too strongly.

\begin{figure}[tb]
\includegraphics[height=.45\textwidth]{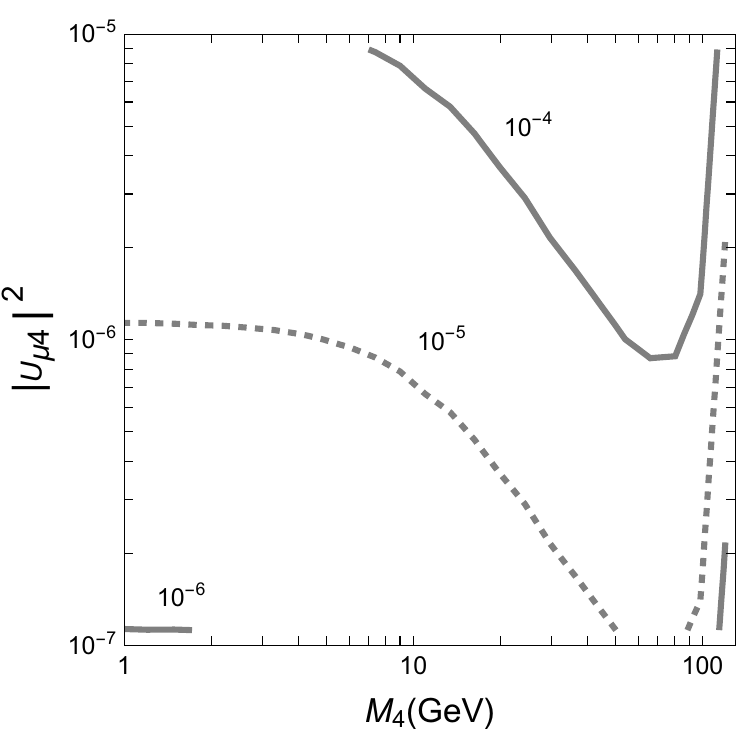}\quad
\includegraphics[height=.45\textwidth]{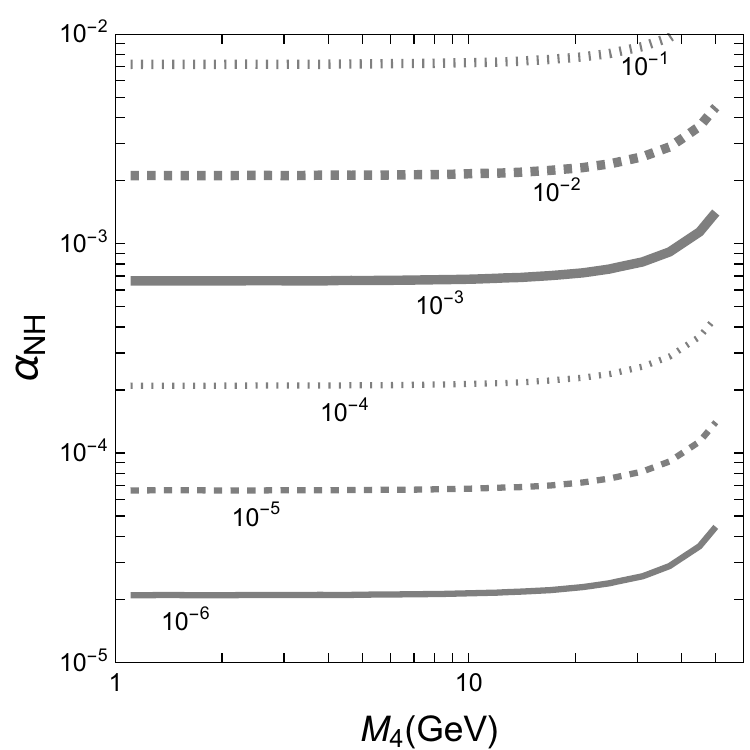}
\caption{Contours for heavy neutrino production cross section, in pb. The left (right) panel shows single (double) heavy neutrino production, for the minimal (extended) models.}
 \label{fig:crosssec}
\end{figure}

The cross section for heavy neutrino production via the decay of a VBF Higgs or $Z$ boson is shown on Figure~\ref{fig:crosssec}, for both minimal and extended models. The left panel shows contours of $\sigma(p p\to H j j)\times{\rm BR}(H\to \nu_\ell\, N_{4,5})+\sigma(p p\to Z j j)\times{\rm BR}(Z\to\nu_\ell\, N_{4,5})$, for the minimal model. Here we sum over all light neutrinos, and over the two heavy neutrinos. The $Z$ contributions are dominant for $M_4\lesssim 10$ GeV, with the branching ratio proportional to $|U_{ah}|^2$. For larger masses the Higgs contributions are more significant, as the couplings increase with $M_4$ (for example, see~\cite{Pilaftsis:1991ug}). The small cross sections are due to the considerably low branching ratios, between $\ord{10^{-9}}$ to $\ord{10^{-3}}$.

The right panel of Figure~\ref{fig:crosssec} shows the cross section for two heavy neutrino production via the decay of a VBF Higgs, $\sigma(p p\rightarrow H j j)\times{\rm BR}(H\rightarrow N_4 N_4)$. Its primary dependence is on the effective parameter of the theory $\alpha_{NH}$, which allows for much larger production rates than the minimal model. As we shall see, this could be a gateway to test $|U_{\mu4}|^2\lesssim10^{-7}$. In the following, for definiteness, we assume that only $(\alpha_{NH})_{44}$ is different from zero.

\begin{figure}[tb]
\includegraphics[height=.35\textwidth]{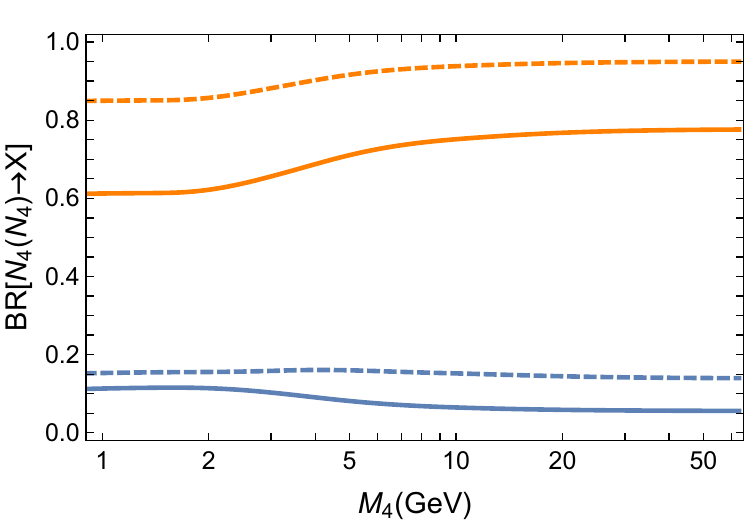}
\caption{Solid lines indicate the branching ratios of heavy neutrino into final states contributing to displaced dileptons (blue) and displaced jets (red). Dashed lines refer to the extended model, and show the probability for the produced heavy neutrino pair to decay into the latter final states.}
 \label{fig:BRs}
\end{figure}
Previous new physics studies at the LHC have analyzed possible cuts on final states of long-lived particles. In the following, we describe cuts from searches for displaced dileptons~\cite{Khachatryan:2014mea, CMS:2016isf} and displaced multitrack jets~\cite{Aaboud:2017iio}. The heavy neutrino branching ratios into final states contributing to each search are shown in Figure~\ref{fig:BRs}. On the same Figure, for the extended model, we also show the probability of having the heavy neutrino pair decaying into such final states. Notice that a same final state can contribute to more than one search, so these branching ratios do not have to add to unity. The probabilities for the extended model are larger than the single heavy neutrino branching ratios since each of the two heavy neutrinos can contribute to the final states.

\subsection{Displaced Dileptons}
\label{sec:dileptonsearch}

This search seeks a displaced, oppositely charged, muon-electron pair. This final state is interpreted as the product of the decay of neutral long-lived particles decaying within the tracker. Having very low background, the search has been used to constrain models of displaced supersymmetry~\cite{Khachatryan:2014mea}.

\begin{table}[tb]
\centering
\setlength{\tabcolsep}{2em}
{\begin{tabular}{| c | c || c | c |} 
\hline
$p_T(e)$ & $>10\,$ GeV & $\Delta R$($\mu,\,e$)& $>0.5$ \\
\hline
$p_T(\mu)$ & $>8\,$ GeV & $\sqrt{L_x^2+L_y^2}$ & $<40\,$mm \\
\hline
$|\,\eta(\ell)\,|$& $<2.4$ & $L_z$ & $<300\,$mm \\
\hline
\end{tabular}}
\caption{Basic cuts for displaced dilepton search.}
\label{table:4}
\end{table}
The specific cuts we use for this channel are shown in Table~\ref{table:4}. The search requires central, isolated, high-$p_T$ leptons with considerable angular separation $\Delta {\rm R} = \sqrt{\Delta\phi^2 + \Delta\eta^2 }$. In addition, the pair is required to be reconstructed within the tracker, implying constraints on $\sqrt{L_x^2+L_y^2}$ and $L_z$, with $L_x$, $L_y$, $L_z$ being the distances in $x$, $y$, $z$ travelled by the heavy neutrino before decaying. The kinematical variable related to the displaced decay is the transverse impact parameter of the leptons, defined as the distance of closest approach to the beamline: 
\begin{ceqn}
\begin{align}
\label{eq:impar}
d_{0} =\dfrac{|p_x^\ell\, L_y-p_y^\ell\, L_x|}{p_T^\ell}
\end{align}
\end{ceqn} 
Here $p^\ell_{x,y}$ are the $x$ and $y$ components of the lepton transverse momentum, $p_T^\ell$. The 13 TeV CMS analysis defines three non-overlapping signal regions (SR):
\begin{itemize}
\item\textbf{SR III}: $|d_0|_{e,\mu} > 1000\, \mu$m
\item\textbf{SR II}: $|d_0|_{e,\mu} > 500\,\mu$m and at least one of the leptons outside of \textbf{SR III}.
\item\textbf{SR I}: $|d_0|_{e,\mu} > 200\,\mu$m and at least one of the leptons outside of \textbf{SR II}.
\end{itemize}
These cuts greatly reduce the SM background. The analysis in~\cite{CMS:2016isf}, for 13 TeV and $2.6\,$fb$^{-1}$, expects less than $3.2$, $0.5$ and $0.02$ background events for \textbf{SR I}, \textbf{II} and \textbf{III}, respectively.

This search was originally triggered by high $p_T$ leptons. In particular, 42 GeV (40 GeV) cuts were applied on the outgoing electron (muon). Since we are triggering the search using VBF, we propose to lower the lepton $p_T$ cuts. We display the minimum cuts in Table~\ref{table:4}, and discuss other possibilities in Section~\ref{sec:dileptonresults}.

Note that a similar search was carried out in~\cite{Caputo:2017pit} for the extended model, using gluon fusion and standard charged lepton triggers, based on~\cite{CMS:2015pca,CMS:2014hka}. A direct comparison between the latter and our work is difficult, as each search considers different cuts. In particular, \cite{CMS:2015pca,CMS:2014hka} use a trigger requiring lower values of $p_T$. We expect that using a higher $p_T$, as is done in~\cite{CMS:2016isf}, would significantly reduce the sensitivity of~\cite{Caputo:2017pit}.

\subsection{Displaced Jets}
\label{sec:displacedjetsearch}

This search is the hadronic analogue of the displaced dilepton search, targeting neutral long-lived particles decaying into jets within the tracker~\cite{Aaboud:2017iio}. Information from the latter is used\footnote{Apart from the standard tracking algorithms, the search includes a \textit{large-radius tracking} algorithm, with relaxed requirements on the longitudinal $z_0$ and transverse $d_0$ impact parameters.} to reconstruct both primary (PV) and displaced vertices (DV). In order to identify a DV, the search requires it to have at least 5 associated tracks, which must be charged and stable, with $p_T > 1$ GeV and $|d_0| > 2$ mm. These must reconstruct at least one \textit{trackless jet}\footnote{Trackless jets are defined as jets for which $\sum p^{track}_T<5 $ GeV, where the sum is taken over all tracks.} with $p_T > 70$ GeV, or two trackless jets with $p_T > 25$ GeV. Further cuts are imposed on the position of the DV within the detector, as well as on the reconstructed invariant mass, $m_{DV}$. The cuts we use, taken from~\cite{Aaboud:2017iio}, are summarized in Table~\ref{table:3}. Once the cuts are applied, one needs to take into account the DV reconstruction efficiency, which is a function of the vertex distance, number of tracks and $m_{DV}$. This is also included in our analysis, and taken from auxiliary material in~\cite{Aaboud:2017iio}.

\begin{table}[tb]
\centering
\setlength{\tabcolsep}{1.2em}
{\begin{tabular}{| c | c || c | c |} 
  \hline
 $p_T(j_1,\,j_2)$ &$>(70,\,0)\,$GeV $\vee$ $>(25,\,25)\,$GeV & $|d_0|$ & $>2\,$mm \\
  \hline
 $|\eta(j_i)|$ & $<$ 4.9 & $\left(\sqrt{L_x^2+L_y^2}\right)_{\rm min}$ & 4 mm.\\
  \hline
 ${\#}$ of tracks & $\geq5$ & $\left(\sqrt{L_x^2+L_y^2}\right)_{\rm max}$ & 300 mm.\\
 \hline
  $m_{DV}$ & $>10\,$GeV & $|L_z|$ & $<300\,$mm.\\
 \hline
\end{tabular}}
 \caption{Cuts for displaced jet searches.}
\label{table:3}
\end{table}

The original search involves a $p_T^{\rm miss}>200\,$GeV trigger. We opt to change this in favour of using VBF. Similar cases have been studied recently for a left-right extension of the minimal model, in which the missing energy trigger is replaced by a prompt lepton trigger~\cite{Cottin:2018kmq,Cottin:2019drg}.

\section{Results}
\label{sec:results}

In this Section we simulate heavy neutrino production through VBF, and evaluate the two searches presented above. For the minimal model, production happens through both Higgs and $Z$ boson mediators, while for the extended model we consider production exclusively by the Higgs.

\subsection{Displaced Dileptons}
\label{sec:dileptonresults}

For the minimal model, the displaced pair is produced via the decay $N_{4,5}\to\nu_\ell\, e^{\pm}\,\mu^{\mp}$. As shown in Figure~\ref{fig:BRs}, the branching ratio into this channel is very small, regardless of our choice for the parametrization. On the effective model the branching ratio is somewhat larger, since we also have the possibility of each heavy neutrino decaying into states with only one lepton, such as $N_4\to\ell\,j\,j$.

\begin{table}[p]
\centering
\setlength{\tabcolsep}{2em}
{\begin{tabular}{| c | c | c|  c|}
\hline
Criterion & $|U_{\mu4}|^2=10^{-6}$  &  $|U_{\mu4}|^2=10^{-7}$& $|U_{\mu4}|^2=10^{-7}$\\
&$M_4=14$GeV&$M_4=14$GeV&$M_4=50$GeV\\
\hline
VBF & 5767 & 5457& 6101\\
\hline
Contained& 5760 & 5060& 6100\\
\hline
Isolated& 2893 & 2532& 5185\\
\hline
$|\,\eta(e,\mu)\,|<2.4$ & 2424 & 2216& 4611\\
\hline
$\Delta R(\mu,\,e)>0.5$ &1933 & 1735& 4119\\
\hline
SR1 & 53/4/0 & 61/24/0&1/0/0 \\
\hline
SR2 & 7/6/0 & 121/47/9& 0/0/0\\
\hline
SR3 & 1/0/0 & 65/29/7& 1/0/0\\
\hline
\end{tabular}}
\caption{Cut flow  for the displaced dilepton search in the minimal model, for several benchmarks. We have $10^5$ initial events. \textit{Contained} events are those where the heavy neutrino decays within the tracker. Each signal region shows three numbers, corresponding to the number of events after the loose, medium and tight cuts on the charged lepton $p_T$. Events on each point have been generated with different seeds.}
\label{table:cutflow.dilep.min}
\end{table}

\begin{table}[p]
\centering
\setlength{\tabcolsep}{1.5em}
{\begin{tabular}{| c | c | c|  c|}
\hline
Criterion & $|U_{\mu4}|^2=10^{-7}$  &  $|U_{\mu4}|^2=10^{-9}$& $|U_{\mu4}|^2=10^{-11}$\\
&$M_4=14$GeV&$M_4=25$GeV&$M_4=50$GeV\\
\hline
VBF & 12643 & 12769& 12621\\
\hline
Contained& 11444 & 7539& 6803\\
\hline
Isolated&  2278& 4260& 5657
\\
\hline
$|\,\eta(e,\mu)\,|<2.4$ & 1826 & 3356& 4886\\
\hline
$\Delta R(\mu,\,e)>0.5$ &1704 & 3098& 4512\\
\hline
SR1 & 202/106/19 & 232/155/37&271/47/23 \\
\hline
SR2 & 130/33/13 & 155/52/26& 514/94/47\\
\hline
SR3 & 104/17/3 & 697/181/15& 1801/561/187\\
\hline
LHC events ($\alpha_{NH}=10^{-3}$)&0/0/0 &  2/0/0& 2/1/0\\
\hline
LHC events ($\alpha_{NH}=10^{-2}$)&52/18/4 &  151/54/11& 183/50/18\\
\hline
\end{tabular}}
\caption{As in Table~\ref{table:cutflow.dilep.min}, but for the extended model. On the last two rows we show the number of events expected at the LHC for two values of $\alpha_{NH}$.}
\label{table:cutflow.dilep.ex}
\end{table}

\afterpage{\clearpage}

For both models, we apply the cuts described in Section~\ref{sec:dileptonsearch} on the two most energetic, oppositely charged, $e-\mu$ pair. As mentioned, we will relax the $p_T$ cut on the charged leptons, evaluating instead the impact of imposing \textit{loose}, \textit{medium} and \textit{tight} cuts. For electrons (muons), loose cuts are of $p_T>10$ ($8$) GeV. Medium and tight cuts mean $p_T>20$ and $30$ GeV, respectively, for both kinds of charged lepton. Since we do not have a way to properly estimate the background nor the detector efficiency, the analysis is performed over the truth level events, with no detector simulation. Results from the cutflow are reported in Tables~\ref{table:cutflow.dilep.min} and~\ref{table:cutflow.dilep.ex}, for different points on the parameter space. On each signal region, we report the number of events surviving the three $p_T$ cuts.

As mentioned earlier, for the minimal model the production cross-section is proportional to $|U_{a4}|^2$, meaning that the mixing cannot be too small. This in turn means that the heavy neutrino cannot be too heavy, or its lifetime will be too short to leave a displaced signal. 
Given the relatively low mass of $N_4$, the final state leptons show $p_T$ distributions favouring low values of momenta. The high $p_T$ leptons surviving the cuts come from highly boosted $N_4$, which lead to collimated $e-\mu$ pairs with a very small $\Delta R$. This is disfavoured by the search, which requires isolated leptons with $\Delta R>0.5$, and is reflected in the cutflow of Table~\ref{table:cutflow.dilep.min}. Even in the best case scenario, shown on the middle column, we find that once the integrated luminosity, cross-section and branching ratios are convoluted, this scenario is not observable, even at the HL-LHC.

We see a significant improvement in the extended model. Here, the new production process gives us the freedom to decrease the mixing down to the seesaw limit without affecting the cross-section, meaning that masses up to $\sim60\,$GeV can be probed. The lepton pair can also have larger $\Delta R$ distances, since they will usually originate from different heavy neutrinos. The difficulties with this scenario are the isolation cuts, as recognized in~\cite{Liu:2019ayx}, and having the leptons pass the $p_T$ cuts. This occurs because the Higgs decays this time into two heavy neutrinos, so the available boost factor for the leptons is less than in the minimal model.  Nevertheless, the situation is much better than in the previous case, and one does obtain a relatively large number of events surviving the cuts at the LHC.

An additional feature of the extended model in this search is that, since now the heavy neutrino can have hadronic decays, each event is more likely to pass the $\sum_j p_T$ cut of the VBF trigger.

\begin{figure}[tb]
\includegraphics[height=.41\textwidth]{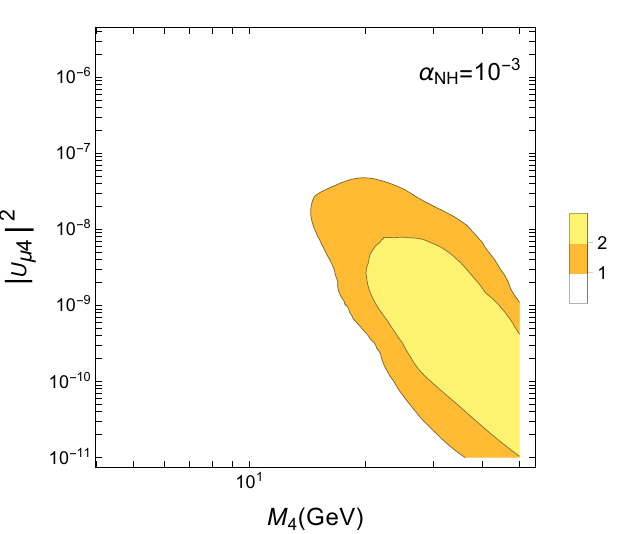}\quad
\includegraphics[height=.41\textwidth]{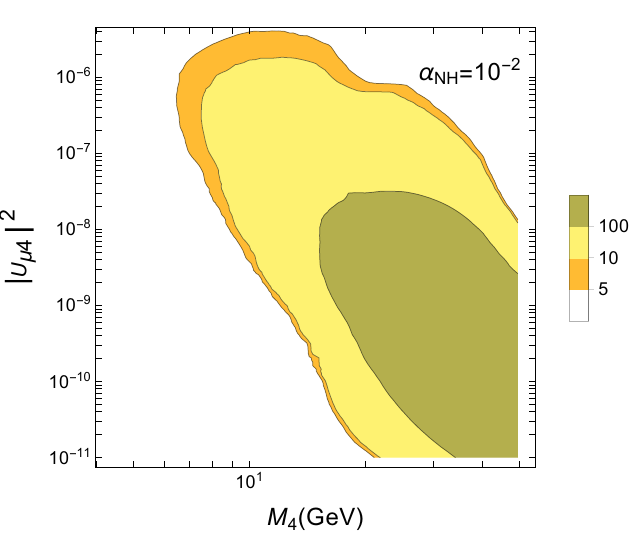}
\caption{Number of displaced dilepton events at the LHC, for the extended model. Only loose $p_T$ cuts are applied.}
 \label{fig:nevents}
\end{figure}
The expected number of events for the extended model is shown in Figure~\ref{fig:nevents}. We apply loose $p_T$ cuts, and sum over the three signal regions. Results for $\alpha_{NH}=10^{-3}$ and $\alpha_{NH}=10^{-2}$ are presented on the left and right panels, respectively. We do not consider larger values of $\alpha_{NH}$ in order to stay below the current limit of $24\%$ for the Higgs to invisible decay~\cite{Tanabashi:2018oca}. We take $M_4$ up to 60 GeV, since the Higgs would not decay to two heavy neutrinos for larger masses, and $|U_{\mu4}|^2$ down to $10^{-11}$, which is close to the seesaw limit.

On the Figure, we find that it is possible to have more than 10 events only for values of $\alpha_{NH}$ close to the current bound. Contrary to the typical plots for displaced vertices, we obtain a larger number of events for very large masses and very small mixing. This happens because, as mentioned before, in this model the production cross-section does not depend on $|U_{a4}|^2$. Thus, one can probe masses as large as those allowed by kinematics, as long as the mixing keeps the heavy neutrino long lived. In our case, larger masses lead to final states with larger values of $p_T$, which pass the cuts more easily.

In this Section we only quote signal yields because the expected background events are mainly coming from instrumental sources. In general this type of background is very difficult to simulate accurately with usual MC tools. Even within the experimental collaborations, data-driven techniques are used in order to get a realistic background prediction. Therefore, we cannot get a real expectation for the sensitivity of this search for our signals. Nevertheless we do expect to have a significant decrease of backgrounds by including the VBF criteria along with the displaced dileptons selection.

Assuming a discovery in the extended model, it would be desirable to verify the existance of exactly two displaced vertices per event. For this search, this amounts to requiring that we find two vertices satisfying the $\sqrt{L_x^2+L_x^2}$ and $L_z$ constraints. For $\alpha_{NH}=10^{-2}$, on the first two benchmarks of Table~\ref{table:cutflow.dilep.ex}, this extra requirement reduces the number of events by a factor $\sim0.8$ for the loose and medium cuts, and $\sim0.6$ for the tight cut. The events for the third benchmark are reduced by a factor $\sim0.5$ for the loose and medium cuts, while for the tight cuts the factor is $0.06$.

\subsection{Displaced Jets}

For this search, any heavy neutrino decay involving quarks can lead to a displaced jet signal, for both models. This is reflected on the much larger branching ratio shown in Figure~\ref{fig:BRs}.

We use as base framework the Recasting Codes for Displaced Vertex Repository~\cite{Brooijmans:2018xbu}, which takes as input the LHE files generated via \texttt{MadGraph}. In the following, we use the detector simulation of~\cite{Allanach:2016pam}, based on \texttt{PYTHIA} and \texttt{FASTJET 3.3.2}~\cite{Cacciari:2011ma}, to impose the displaced jet cuts over the events. The decay position for the heavy neutrino is obtained via \texttt{MadGraph} with the time of flight option. This method has been used in~\cite{Cottin:2018kmq,Cottin:2019drg} for the same final state, but with a different trigger process.

\begin{table}[p]
\centering
\setlength{\tabcolsep}{2em}
{\begin{tabular}{| c | c | c|  c|} 
\hline
Criterion & $|U_{\mu4}|^2=10^{-6}$  &  $|U_{\mu4}|^2=10^{-7}$& $|U_{\mu4}|^2=10^{-7}$\\
&$M_4=14$GeV&$M_4=14$GeV&$M_4=50$GeV\\
\hline
VBF & 21462 & 21494& 20506\\
\hline
Trackless Jets& 21442 & 21469 &20499\\
\hline
Contained & 21440 & 21315&20499 \\
\hline
Displaced tracks &28 & 1021& 0\\
\hline
$m_{DV}>10\,$GeV &5 & 85&0 \\
\hline
DV efficiency & 0 & 11& 0\\
\hline
\end{tabular}}
\caption{Cut flow  for the displaced jet search in the minimal model, for several benchmarks. We have $10^5$ initial events. \textit{Contained} events are those where the heavy neutrino decays within the tracker. The events surviving the \textit{displaced tracks} cut are those with at least five tracks with $d_0>2\,$mm and $\sqrt{L_x^2+L_y^2}>4\,$mm.}
\label{table:emucutf1}
\end{table}

\begin{table}[p]
\centering
\setlength{\tabcolsep}{1.5em}
{\begin{tabular}{| c | c | c|  c|} 
\hline
Criterion & $|U_{\mu4}|^2=10^{-7}$  &  $|U_{\mu4}|^2=10^{-9}$& $|U_{\mu4}|^2=10^{-11}$\\
&$M_4=14$GeV&$M_4=25$GeV&$M_4=50$GeV\\
\hline
VBF & 22669 & 22318& 22422\\
\hline
Trackless Jets& 22651 & 22308 &22411\\
\hline
Contained & 22525 & 21368& 21255\\
\hline
Displaced tracks &1446 & 5995& 9853\\
\hline
$m_{DV}>10\,$GeV &72 & 3795&8802 \\
\hline
DV efficiency & 24 & 1016& 3657\\
\hline
LHC events ($\alpha_{NH}=10^{-3}$)&$<$1 &  6& 5\\
\hline
LHC events ($\alpha_{NH}=10^{-2}$)&10 & 577& 487\\
\hline
\end{tabular}}
\caption{As in Table~\ref{table:emucutf1}, but for the extended model. On the last two rows we show the number of events expected at the LHC for two values of $\alpha_{NH}$.}
\label{table:emucutf2}
\end{table}
\afterpage{\clearpage}
The cutflows for this search are shown in Tables~\ref{table:emucutf1} and~\ref{table:emucutf2} for the minimal and extended models, respectively. In both models we find that requiring five displaced tracks leads to a considerable reduction of events. This is in part due to the relatively small $M_4$, since the two quarks from heavy neutrino decay become too soft to generate five tracks from the parton shower\footnote{This was recognized in~\cite{Cottin:2018kmq}, where they suggested reducing the number of displaced tracks to three.}. This situation is worsened by the cut on $m_{DV}$. Thus, we find that the minimal model again cannot be probed by this search, but the extended one, which allows larger masses, survives the cuts provided $\alpha_{NH}$ is large enough.

Similar to the situation for the extended model in the dilepton search, here we find that the hadronic final states in both models contribute to the $\sum_j p_T$ cut of the VBF trigger, leading to a larger number of events being recorded.

Similar to what we did before, we remove the missing energy trigger used in the displaced jets search. In order to calculate the sensitivity of the search to our signal, it is necessary to estimate the expected background at the LHC. According to what is described in~\cite{Aaboud:2017iio}, the only background contribution comes from instrumental sources. Therefore, at the phenomenological level, we do not have a method to simulate such events. However, we can use the information already given by the collaboration in order to get an estimate of the background events when replacing the missing energy criterion by the VBF criteria.

According to their estimate for $32.8\,$fb$^{-1}$, $4.5$ events of background are expected when all displaced jet criteria are applied except for missing energy. To stay on the conservative side, we can use the same VBF criteria efficiency for signal as for background. As our background is only instrumental, we do not expect these events to reflect the VBF topology properties, and therefore the VBF criteria efficiency for background events should be smaller compared to signal efficiency. Thus, using VBF signal efficiency for background events gives us an overestimate of background expectation, which implies that our derived conclusions will be conservative. Real limits should be stronger than our findings.

We calculate the VBF criteria efficiency on signal by producing signal events with \texttt{MadGraph} and applying directly the VBF criteria on the events that passed the displaced jets criteria. We obtain that the efficiency for signal is around $0.07$. Following the strategy described, we then expect at least $0.32$ background events after the full selection, equivalent to around 3 events for an integrated luminosity of 300 fb$^{-1}$.

\begin{figure}[tb]
\includegraphics[height=.41\textwidth]{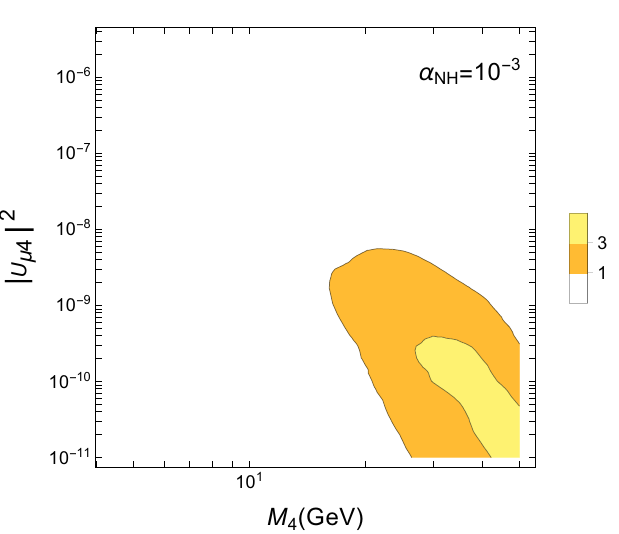}\quad
\includegraphics[height=.41\textwidth]{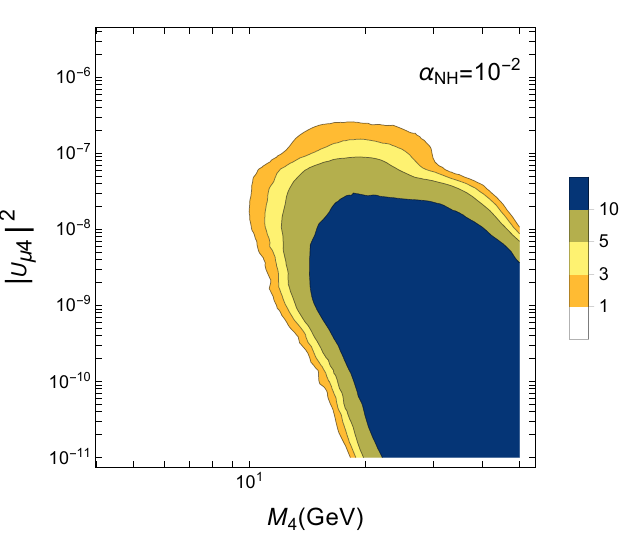}
\caption{Sensitivity of displaced jet search for the extended model at the LHC, for different choices of $\alpha_{NH}$.}
 \label{fig:sensitivity}
\end{figure}
Using the background estimate and scanning over signal we can calculate the expected sensitivity for the extended model. Given the small number of background events, this is calculated using~\cite{Cowan:2010js}:
\begin{equation}
    Z_A=\sqrt{2\left((\text{s}+\text{b})\ln\left(1+\frac{\text{s}}{\text{b}}\right)-\text{s}\right)}
\end{equation}
where $s$ and $b$ are the number of signal and background events, respectively.

Results are shown in Figure~\ref{fig:sensitivity}. As in Figure~\ref{fig:nevents}, the displayed maximum mass and minimum mixing are bounded by kinematical and seesaw limits, respectively. We find that sensitivity is greatest for large values of $M_4$, and accordingly small values of $|U_{\mu4}|^2$. This is due to the larger number of events passing the cuts on those regions, similar to what occurs for the displaced dilepton search. The largest number of signal events we obtain is 847, for $M_4\sim40$~GeV and $|U_{\mu4}|^2\sim10^{-9}$. For even larger masses the number of events decreases due to a smaller production cross-section, as shown on the right panel of Figure~\ref{fig:crosssec}.

Remarkably, for the extended model, we find that the displaced jet search triggered by VBF can have more than $5\sigma$ sensitivity to certain regions of the parameter space.  Even for $\alpha_{NH}=10^{-3}$, it is feasible to acquire evidence in favour of this coupling between the Higgs and heavy neutrinos.

As was done for the dilepton case, we have assessed the possibility of measuring exactly two displaced vertices per event. On the three benchmark points shown in Table\ref{table:emucutf2}, we find that, for $\alpha_{NH} = 10^{-2}$, the combination of displaced track, $m_{DV}$ and DV efficiency cuts reduce the total number of events to zero, four and eight for each benchmark. Thus, a two-vertex search is unfortunately not feasible in this channel, at the LHC.

\section{Conclusions}
\label{sec:conclusions}

In this work we propose using a VBF trigger in long-lived heavy neutrino  production. We have explored a minimal GeV-scale seesaw model, as well as an extension with an effective operator. On the minimal model, VBF production is dominated either by an intermediate $Z$ or Higgs boson decaying into one light and one heavy neutrino. On the extended one, production occurs exclusively through Higgs decay into two heavy neutrinos.

After production, we have considered two different searches: displaced dileptons and displaced multitrack jets. On each search, we have dropped their proposed trigger, and replaced it by VBF. We find that the minimal model does not survive the combination of cuts, suggesting that other searches might be more appropriate. For the extended model, we have demonstrated that both searches can be used. Moreover, for displaced multitrack searches we have shown that, in spite of a conservative background estimate, we can achieve a sensitivity larger than $5\sigma$ if the effective coupling is large enough. This allows us to explore regions with large masses and very small mixing, which are hard to reach in $W^\pm$-mediated production.

Thus, we can conclude that the VBF trigger can be used to scrutinize the neutral current couplings of long-lived heavy neutrinos, being complementary to triggers involving prompt leptons, sensitive to charged currents.

\section*{Acknowledgements}

J.J.P. and J.~M.~acknowledge funding by the {\it Direcci\'on de Gesti\'on de la Investigaci\'on} at PUCP, through grant DGI-2019-3-0044. J.~M.~was also funded by the {\it Programa de Apoyo al Desarrollo de Tesis de Licenciatura} (PADET). J.~D.~R.-Á. gratefully acknowledges the support of the Colombian Science Ministry MinCiencias and Sostenibilidad-UdeA.

\appendix

\section{Parametrisation}

In this work we follow the parametrisation of~\cite{Donini:2012tt}, which for our purposes is identical to the one of Casas-Ibarra~\cite{Casas:2001sr}. The $6\times6$ mixing matrix $U$ is decomposed into four $3\times3$ blocks:
\begin{equation}
 U=\left(\begin{array}{cc}
U_{a\ell} & U_{ah} \\
U_{s\ell} & U_{sh}
\end{array}\right)~.
\end{equation}
Each block can be parametrised in the following way:
\begin{align}
 \label{eq:mixingmats}
 U_{a\ell} &= U_{\rm PMNS}\,H~,  & 
 U_{ah} &= i\,U_{\rm PMNS}\,
 H\,m_\ell^{1/2}R^\dagger M_h^{-1/2}~, \nonumber \\
 U_{s\ell} &= i\, \bar W\bar H M_h^{-1/2}\,R\,m_\ell^{1/2}~, &
 U_{sh} &= \bar W\bar H~,
\end{align}
In the Equations above, the unitary matrix $U_{PMNS}$ would correspond to the observed PMNS neutrino mixing matrix on the limit $H\to I$. We use oscillation parameters as those found in~\cite{deSalas:2017kay,Esteban:2018azc}. The $\bar W$ matrix is also unitary, and allows us to redefine the basis of the sterile neutrinos, in this parametrisation it is set equal to the identity. Departures of unitarity in the active-light and sterile-heavy sectors are encoded on the $H$ and $\bar H$ matrices:
\begin{eqnarray}
\label{eq:hreal}
H &=& \left(I+m_\ell^{1/2}\,R^\dagger\,M_h^{-1}\,R\,m_\ell^{1/2}\right)^{-1/2} \\
\label{eq:hbarreal}
\bar H &=& \left(I+M_h^{-1/2}\,R\,m_\ell\,R^\dagger\,M_h^{-1/2}\right)^{-1/2}~,
\end{eqnarray}
These can be approximated $H\approx\bar H\approx I$ in the region of our interest.

The diagonal matrix $m_\ell$ contains the light neutrino masses, and depends on the ordering (normal NO or inverted IO):
\begin{align}
m_\ell&={\rm diag}(m_1,\,m_2,\,m_3)={\rm diag}\left(m_1,\,\sqrt{\Delta m^2_{\rm sol}+m_1^2},\sqrt{\Delta m^2_{\rm atm}+m_1^2}\right)&{\rm (NO)} \\
m_\ell&={\rm diag}(m_3,\,m_1,\,m_2)={\rm diag}\left(m_3,\,\sqrt{|\Delta m^2_{\rm atm}|-\Delta m^2_{\rm sol}+m_3^2},\sqrt{|\Delta m^2_{\rm atm}|+m_3^2}\right)&{\rm (IO)} 
\end{align}
where $\Delta m^2_{\rm sol}$ and $\Delta m^2_{\rm atm}$ are the squared mass differences measured in solar and atmospheric neutrino oscillation experiments, respectively. The diagonal matrix $M_h={\rm diag}(M_4,\,M_5,\,M_6)$ contains the heavy neutrino masses, with no order in particular.

The crucial part of the parametrisation lies on the complex orthogonal matrix $R$, which allows an enhancement to the active-heavy mixing. We write:
\begin{equation}
 R=\left(\begin{array}{ccc}
 \tilde c_{45} & \tilde s_{45} & 0\\
 -\tilde s_{45} & \tilde c_{45} & 0 \\
 0 & 0 & 1
\end{array}\right)
\left(\begin{array}{ccc}
 \tilde c_{46} & 0 & -\tilde s_{46} \\
 0 & 1 & 0 \\
 \tilde s_{46} & 0 & \tilde c_{46} \\
\end{array}\right)
\left(\begin{array}{ccc}
1 & 0 & 0 \\
0 & \tilde c_{56} & \tilde s_{56}\\
0 & -\tilde s_{56} & \tilde c_{56}
\end{array}\right)
~.
\end{equation}
where $\tilde s_{ij}$ and $\tilde c_{ij}$ are the sines and cosines of a complex angle, $\theta_{ij}+i\gamma_{ij}$. It is useful to expand the complex angles in the $R$ matrix so, for example:
\begin{multline}
R_{45}(\theta_{45},\gamma_{45})\equiv \begin{pmatrix}
\tilde c_{45} & \tilde s_{45} & 0 \\
-\tilde s_{45} & \tilde c_{45} & 0 \\
0 & 0 & 1
 \end{pmatrix} \\
 =\left(\begin{array}{ccc}
 c_{45}\cosh\gamma_{45}-i\,s_{45}\sinh\gamma_{45} & s_{45}\cosh\gamma_{45}+i\,c_{45}\sinh\gamma_{45} & 0\\
-s_{45}\cosh\gamma_{45}-i\,c_{45}\sinh\gamma_{45} & c_{45}\cosh\gamma_{45}-i\,s_{45}\sinh\gamma_{45} & 0 \\
0 & 0 & 1
\end{array}\right)~,
\end{multline}
where $s_{ij}$ and $c_{ij}$ are the sine and cosine of the real angle $\theta_{ij}$. For large $|\gamma_{45}|\gtrsim3$, we have $|\sinh\gamma_{45}|\approx\cosh\gamma_{45}$, such that $\theta_{45}$ behaves as an overall phase:
\begin{equation}
 R_{45}(\theta_{45},{|\gamma_{45}|\gg0})=\left(\begin{array}{ccc}
\cosh\gamma_{45}\,e^{- i z_{45}\,\theta_{45}} & i\,z_{45}\cosh\gamma_{45}\,e^{-i z_{45}\,\theta_{45}} & 0  \\
- i\,z_{45}\cosh\gamma_{45}\,e^{-i z_{45}\,\theta_{45}} & \cosh\gamma_{45}\,e^{-iz_{45}\,\theta_{45}} & 0 \\
0 & 0 & 1
\end{array}\right)~,
\end{equation}
where $z_{45}$ is the sign of $\gamma_{45}$. This allows for simplified ways of writing down the elements of the active-heavy sector $U_{ah}$.

In the next subsections, we write compact expressions using the latter approximation. We always assume $H\sim I$, and denote $z_{ij}=\pm1$ as the signs of $\gamma_{ij}$. One finds that, in general, two heavy neutrinos have their mixing enhanced by at most three powers of $\cosh\gamma_{ij}$, with the third one having one lesser power. Notable exceptions are cases~\ref{item:allenhanced1}, \ref{item:allenhanced2} and \ref{item:allenhanced3}, where all three heavy neutrinos can have the same powers of $\cosh\gamma_{ij}$, depending on the choice of mixing angles $\theta_{mn}$. In the following, we only write leading terms, that is, those with larger powers of $\cosh\gamma_{ij}$.

We present results valid for normal ordering. In this case, we find that the mixing of heavy neutrinos with the active electron state are generally smaller by an order of magnitude, due to a $|(U_{\rm PMNS})_{a3}|=s_{13}$ suppression. This is avoided explicitly in~\ref{item:No-s13-1}, where the mixing is controlled mainly by $|(U_{\rm PMNS})_{a2}|=c_{13}\,s_{12}$, and can be further enhanced by taking values of $m_1$ close to the experimental limit. A similar situation can be reproduced in~\ref{item:No-s13-2}, \ref{item:allenhanced1} and \ref{item:No-s13-3} through specific choices of $\theta_{ij}$.

\begin{enumerate}[label=(\roman*)]
    
\item Non-zero $\gamma_{45}$, $\theta_{45}$ :
\label{item:No-s13-1}
\begin{eqnarray}
U_{a4}
&=& z_{45}\,Z_a\,\sqrt{\frac{m_2}{M_4}}\cosh\gamma_{45}\,e^{i\,z_{45}\,\theta_{45}} \\
U_{a5}
&=& i\,Z_a\sqrt{\frac{m_2}{M_5}}\cosh\gamma_{45}\,e^{i\,z_{45}\,\theta_{45}} \\
U_{a6}
&=& i\,(U_{\rm PMNS})_{a3}\sqrt{\frac{m_3}{M_6}} \\
Z_a &=& (U_{\rm PMNS})_{a2}+i\,z_{45}\,\sqrt{\frac{m_1}{m_2}}(U_{\rm PMNS})_{a1}
\end{eqnarray}

\item Non-zero $\gamma_{46}$, $\theta_{46}$ :
\begin{eqnarray}
U_{a4}
&=& -z_{46}\,Z_a\,\sqrt{\frac{m_3}{M_4}}\cosh\gamma_{46}\,e^{i\,z_{46}\,\theta_{46}} \\
U_{a5}
&=& i\,(U_{\rm PMNS})_{a2}\sqrt{\frac{m_2}{M_5}} \\
U_{a6}
&=& i\,Z_a\sqrt{\frac{m_3}{M_6}}\cosh\gamma_{46}\,e^{i\,z_{46}\,\theta_{46}} \\
Z_a &=& (U_{\rm PMNS})_{a3}-i\,z_{46}\,\sqrt{\frac{m_1}{m_3}}(U_{\rm PMNS})_{a1}
\end{eqnarray}

\item Non-zero $\gamma_{56}$, $\theta_{56}$ :
\begin{eqnarray}
U_{a4}
&=& i\,(U_{\rm PMNS})_{a1}\sqrt{\frac{m_1}{M_4}} \\
U_{a5}
&=& z_{56}\,Z_a\sqrt{\frac{m_3}{M_5}}\cosh\gamma_{56}\,e^{i\,z_{56}\,\theta_{56}} \\
U_{a6}
&=& i\,Z_a\,\sqrt{\frac{m_3}{M_6}}\cosh\gamma_{56}\,e^{i\,z_{56}\,\theta_{56}} \\
Z_a &=& (U_{\rm PMNS})_{a3}+i\,z_{56}\,\sqrt{\frac{m_2}{m_3}}(U_{\rm PMNS})_{a2}
\end{eqnarray}

\item Non-zero $\gamma_{45}$, $\theta_{ij}$ :
\label{item:No-s13-2}
\begin{eqnarray}
U_{a4}
&=& z_{45}\,Z_a\,\sqrt{\frac{m_3}{M_4}}\cosh\gamma_{45}\,e^{i\,z_{45}\,\theta_{45}} \\
U_{a5}
&=& i\,Z_a\,\sqrt{\frac{m_3}{M_5}}\cosh\gamma_{45}\,e^{i\,z_{45}\,\theta_{45}} \\
U_{a6}
&=& i\,Y_a\,\sqrt{\frac{m_3}{M_6}} \\
Z_a &=& (U_{\rm PMNS})_{a3}(s_{56}-i\,z_{45}\,s_{46}c_{56})+\sqrt{\frac{m_2}{m_3}}(U_{\rm PMNS})_{a2}(c_{56}+i\,z_{45}\,s_{46}s_{56}) \nonumber \\
&&\qquad+i\,z_{45}\,\sqrt{\frac{m_1}{m_3}}(U_{\rm PMNS})_{a1}\,c_{46} \\
Y_a &=& (U_{\rm PMNS})_{a3}\,c_{46}c_{56}-\sqrt{\frac{m_2}{m_3}}(U_{\rm PMNS})_{a2}\,c_{46}s_{56}
+\sqrt{\frac{m_1}{m_3}}(U_{\rm PMNS})_{a1}\,s_{46}
\end{eqnarray}

\item Non-zero $\gamma_{46}$, $\theta_{ij}$ :
\label{item:allenhanced1}
\begin{eqnarray}
U_{a4}
&=& -z_{46}\,Z_a c_{45}\,\sqrt{\frac{m_3}{M_4}}\cosh\gamma_{46}\,e^{i\,z_{46}\,\theta_{46}} \\
U_{a5}
&=& z_{46}\,Z_a s_{45}\,\sqrt{\frac{m_3}{M_5}}\cosh\gamma_{46}\,e^{i\,z_{46}\,\theta_{46}} \\
U_{a6}
&=& i\,Z_a\,\sqrt{\frac{m_3}{M_6}}\cosh\gamma_{46}\,e^{i\,z_{46}\,\theta_{46}} \\
Z_a &=& (U_{\rm PMNS})_{a3}\,c_{56}-\sqrt{\frac{m_2}{m_3}}(U_{\rm PMNS})_{a2}\,s_{56} -i\,z_{46}\,\sqrt{\frac{m_1}{m_3}}(U_{\rm PMNS})_{a1}
\end{eqnarray}

\item Non-zero $\gamma_{56}$, $\theta_{ij}$ :
\label{item:allenhanced2}
\begin{eqnarray}
U_{a4}
&=& -i\,Z_a\,(c_{45}s_{46}+i\,z_{56}\,s_{45})\,\sqrt{\frac{m_3}{M_4}}\cosh\gamma_{56}\,e^{i\,z_{56}\,\theta_{56}} \\
U_{a5}
&=& i\,Z_a\,(s_{45}s_{46}-i\,z_{56}\,c_{45})\,\sqrt{\frac{m_3}{M_5}}\cosh\gamma_{56}\,e^{i\,z_{56}\,\theta_{56}} \\
U_{a6}
&=& i\,Z_a c_{46}\,\sqrt{\frac{m_3}{M_6}}\cosh\gamma_{56}\,e^{i\,z_{56}\,\theta_{56}} \\
Z_a &=& (U_{\rm PMNS})_{a3}+i\,z_{56}\sqrt{\frac{m_2}{m_3}}(U_{\rm PMNS})_{a2}
\end{eqnarray}

\item Non-zero $\gamma_{45}$, $\gamma_{46}$, $\theta_{ij}$:
\label{item:No-s13-3}
\begin{eqnarray}
U_{a4}
&=& -z_{46}\,Z_a\,\sqrt{\frac{m_3}{M_4}}\cosh\gamma_{45}\cosh\gamma_{46}\,e^{i(z_{45}\,\theta_{45}+z_{46}\,\theta_{46})} \\
U_{a5}
&=& -i\,z_{45}\,z_{46}\,Z_a\,\sqrt{\frac{m_3}{M_5}}\cosh\gamma_{45}\cosh\gamma_{46}\,e^{i(z_{45}\,\theta_{45}+z_{46}\,\theta_{46})} \\
U_{a6}
&=& i\,Z_a\,\sqrt{\frac{m_3}{M_6}}\cosh\gamma_{46}\,\,e^{i\,z_{46}\,\theta_{46}} \\
Z_a &=& (U_{\rm PMNS})_{a3}\,c_{56}-\sqrt{\frac{m_2}{m_3}}(U_{\rm PMNS})_{a2}\,s_{56}-i\,z_{46}\,\sqrt{\frac{m_1}{m_3}}(U_{\rm PMNS})_{a1}
\end{eqnarray}

\item Non-zero $\gamma_{45}$, $\gamma_{56}$, $\theta_{ij}$:
\begin{eqnarray}
U_{a4}
&=& -i\,z_{45}\,Z_a(z_{56}+z_{45}\,s_{46})\sqrt{\frac{m_3}{M_4}}\cosh\gamma_{45}\cosh\gamma_{56}\,e^{i(z_{45}\,\theta_{45}+z_{56}\,\theta_{56})} \\
U_{a5}
&=& Z_a(z_{56}+z_{45}\,s_{46})\sqrt{\frac{m_3}{M_5}}\cosh\gamma_{45}\cosh\gamma_{56}\,e^{i(z_{45}\,\theta_{45}+z_{56}\,\theta_{56})} \\
U_{a6}
&=& i\,Z_a\,c_{46}\,\sqrt{\frac{m_3}{M_6}}\cosh\gamma_{56}\,\,e^{i\,z_{56}\,\theta_{56}} \\
Z_a &=& (U_{\rm PMNS})_{a3}+i\,z_{56}\,\sqrt{\frac{m_2}{m_3}}(U_{\rm PMNS})_{a2}
\end{eqnarray}

\item Non-zero $\gamma_{46}$, $\gamma_{56}$, $\theta_{ij}$:
\label{item:allenhanced3}
\begin{eqnarray}
U_{a4}
&=& -z_{46}\,c_{45}\,Z_a\,\sqrt{\frac{m_3}{M_4}}\cosh\gamma_{46}\cosh\gamma_{56}\,e^{i(z_{46}\,\theta_{46}+z_{56}\,\theta_{56})} \\
U_{a5}
&=& z_{46}\,s_{45}\,Z_a\,\sqrt{\frac{m_3}{M_5}}\cosh\gamma_{46}\cosh\gamma_{56}\,e^{i(z_{46}\,\theta_{46}+z_{56}\,\theta_{56})} \\
U_{a6}
&=& i\,Z_a\,\sqrt{\frac{m_3}{M_6}}\cosh\gamma_{46}\cosh\gamma_{56}\,\,e^{i(z_{46}\,\theta_{46}+z_{56}\,\theta_{56})} \\
Z_a &=& (U_{\rm PMNS})_{a3}+i\,z_{56}\sqrt{\frac{m_2}{m_3}}(U_{\rm PMNS})_{a2}
\end{eqnarray}

\item All $\gamma_{ij}$ large:
\begin{eqnarray}
U_{a4}
&=& -z_{46}\,Z_a\,\sqrt{\frac{m_3}{M_4}}\cosh\gamma_{45}\cosh\gamma_{56}\cosh\gamma_{46}\,e^{i(z_{45}\,\theta_{45}+z_{46}\,\theta_{46}+z_{56}\,\theta_{56})} \\
U_{a5}
&=& -i\,z_{45}\,z_{46}\,Z_a\,\sqrt{\frac{m_3}{M_5}}\cosh\gamma_{45}\cosh\gamma_{56}\cosh\gamma_{46}\,e^{i(z_{45}\,\theta_{45}+z_{46}\,\theta_{46}+z_{56}\,\theta_{56})} \\
U_{a6}
&=& i\,Z_a\,\sqrt{\frac{m_3}{M_6}}\cosh\gamma_{46}\cosh\gamma_{56}\,\,e^{i(z_{46}\,\theta_{46}+z_{56}\,\theta_{56})} \\
Z_a &=& (U_{\rm PMNS})_{a3}+i\,z_{56}\,\sqrt{\frac{m_2}{m_3}}(U_{\rm PMNS})_{a2}
\end{eqnarray}

\end{enumerate}

 One can obtain analogous expressions for inverted ordering by carrying out the following substitutions:
\begin{align}
m_1&\to m_3  & m_2&\to m_1 & m_3&\to m_2 \\
(U_{\rm PMNS})_{a1}&\to (U_{\rm PMNS})_{a3} & (U_{\rm PMNS})_{a2}&\to (U_{\rm PMNS})_{a1} & (U_{\rm PMNS})_{a3}&\to (U_{\rm PMNS})_{a2}
\end{align}

\bibliographystyle{utphys}
\bibliography{vbf}

\begin{thebibliography}{10}
\bibitem{CMS:2015pca}
{\bfseries CMS} Collaboration, ``{Search for long-lived particles that decay
  into final states containing two muons, reconstructed using only the CMS muon
  chambers},''.
\end{thebibliography}

\end{document}